\documentclass[10pt,letterpaper,twocolumn]{article}

\usepackage{ol2}
\usepackage[draft]{hyperref}
\usepackage{amsmath}
\usepackage{epstopdf}
\usepackage{dcolumn}
\usepackage{bm}

\newcommand{\avg}[1]{\langle #1 \rangle}
\newcommand{\abs}[1]{\left|#1\right|}

\begin{document}

\twocolumn[
\title{Non-Imaging Speckle Interferometry for\\High Speed Nanometer-Scale Position Detection}

\author{E.G. van Putten,$^{1,*}$ A. Lagendijk,$^{1,2}$ and A.P. Mosk$^{1}$}

\address{
$^1$Complex Photonic Systems, Faculty of Science and Technology and MESA$^+$ Institute for\\ Nanotechnology, University of Twente, P.O. Box 217, 7500 AE Enschede, The Netherlands \\
$^2$FOM Institute for Atomic and Molecular Physics, Science Park 104, 1098 XG Amsterdam, The
Netherlands\\
$^*$Corresponding author: E.G.vanPutten@alumnus.utwente.nl
}

\begin{abstract}
We experimentally demonstrate a non-imaging approach to displacement measurement for complex scattering materials. By spatially controlling the wave front of the light that incidents on the material
we concentrate the scattered light in a focus on a designated position. This wave front acts
as an unique optical fingerprint that enables precise position detection of the illuminated material
by simply measuring the intensity in the focus.
By combining two optical fingerprints we demonstrate position
detection along one dimension with a displacement resolution of $2.1$~nm.
As our approach does not require an image of the scattered field, it is possible to employ fast non-imaging detectors to enable high-speed position detection of scattering materials.
\end{abstract}

 ]
 
Light is an ideal tool to perform contact free, non-destructive,
and high-precision metrology.\cite{Glasvik2002_book}
For this reason optical positioning techniques have proven themselves
indispensable in various branches of science and
many important industrial processes, including the fabrication of
semi-conductor-based circuits with features on the nanometer regime.
Fast feedback is essential in these high-precision positioning systems
as high frequency vibrational motion limits the level of
precision that these techniques offer.

In smooth reflecting systems, laser interferometry\cite{Bobroff1993_MST} combined with
high speed detectors offers precise displacement measurements with a large bandwidth.
For disordered, rough, or complex surfaces, where conventional laser interferometry
is not applicable,
several speckle based metrology techniques such as speckle
photography\cite{Burch1968_OptAc,Archbold1970_OptAct} and speckle
interferometry\cite{Leendertz1970_JPESI,Butters1971_OLT,Macovski1971_ApplOpt,Lekberg1980_PiT}
were developed
in the late $1960$s and $1970$s.\cite{Goodman2006} These techniques spatially image
speckle patterns to offer versatile measurements of material
parameters such as strain, displacement, and rotation.\cite{Kaufmann2011_book}
However, the necessary spatial information limits the attainable
bandwidth because these techniques require imaging detectors, which
are orders of magnitude slower than non-imaging detectors
such as fast photodiodes, which can have GHz bandwidth.

Recent developments in optics\cite{Vellekoop2007aa,Popoff2010aa,Cizmar2010aa,Choi2011_PRL}
enabled control of the propagation of scattered light.
These techniques, which are conceptually related to
phase conjugation\cite{Leith1966aa,Yaqoob2008aa} and time reversal\cite{Fink2000},
manipulate the wave front of the incident light using spatial light modulators\cite{Maurer2011_LPR} in order to steer the scattered light, for example, in a spatial and/or temporal focus at any desired position\cite{Vellekoop2007aa,Vellekoop2008aa,Aulbach2011aa,Katz2011aa,McCabe2011aa}.

In this Letter we describe and experimentally demonstrate a non-imaging approach to
displacement measurement for complex scattering materials.
We concentrate the light that is scattered from
the material in a sharp focus by spatially shaping the wave front of the incident light.
In a complex system lacking translational invariance, this wave front
acts as an unique optical fingerprint of the illuminated part of the sample.
Any displacement between the fingerprint and the system reduces their overlap,
thereby inevitably decreasing the intensity in the constructed focus. This dependence
opens the possibility to use such optical fingerprints for position detection of the
illuminated sample at resolutions in the order of nanometers. As spatial
information of the scattered field is no longer required, this method furthermore enables the use of fast detectors.

\begin{figure}[htb]
\centerline{\includegraphics[width=7.5cm]{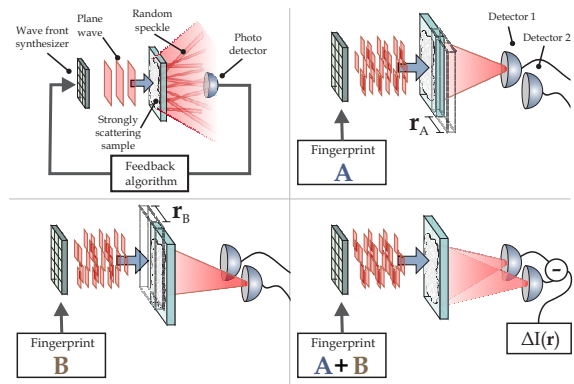}}
\caption{Method to detect sample displacements. Light modulated by a wave front synthesizer
  incidents on a scattering sample. A detector coupled to the wave front synthesizer
  monitors the scattered light. Two optical fingerprints are generated for two sample
  positions. Each of the fingerprints redirects the scattered light onto one of the
  detectors. The sample position $\mathbf{r}$ is then determined by illuminating
  the sample with a superposition of the two fingerprints and monitoring
  the intensity difference $\Delta I$ between the two detectors. While this figure shows
  a horizontal displacement, the method is more generally valid and works in all directions.}
  \label{fig:vanPutten2011_Fig1}
\end{figure}
In Fig.~\ref{fig:vanPutten2011_Fig1} we have depicted our method to detect sample displacements.
With a wave front synthesizer we spatially control
the phase of a light beam. A detector behind
a scattering sample combined with a feedback based algorithm
finds the optical fingerprint that focusses the transmitted light onto the detector.\cite{Vellekoop2008ab}
We use this system to find the fingerprints $A$ and $B$ corresponding to
two different sample positions; $\mathbf{r}_A$ and $\mathbf{r}_B$.
In our current setup, this process takes several minutes. However, very recently it has be shown\cite{Cui2011_OptEx, Choi2011_PRL} that such fingerprints can be found well within a second.
Fingerprint $A$ focusses the light onto the first detector when the sample is positioned at $\mathbf{r}_A$, while fingerprint $B$ is
constructed to concentrate the light onto the second detector when the sample is at $\mathbf{r}_B$. When we now position the sample in between
$\mathbf{r}_A$ and $\mathbf{r}_B$ while illuminating it with a superposition of fingerprints $A$ and $B$, the sample position can be interpolated from the intensity difference of the two
detectors. This method can be easily extended to detect displacements in multiple directions
by generating multiple optical fingerprints at different sample positions.

The sensitivity of this detection method and the corresponding smallest measurable
displacement $\delta \mathbf{r}$ are determined by way the intensities $I_A$
and $I_B$ in the two foci change under a sample displacement. When the sample is illuminated by
an optical fingerprint, the focus intensity $I_0$ as function of the sample displacement
$\Delta \mathbf{r} \equiv \mathbf{r} - \mathbf{r}_0$ from its original position $\mathbf{r}_0$
\begin{align}
    \label{eq:IntensityVsDisplacement}
    I_0(\Delta\mathbf{r}) = \eta \avg{I_\text{bg}}\avg{\abs{\gamma(\Delta\mathbf{r})}^2},
\end{align}
where $\avg{\cdot}$ denotes ensemble averaging over disorder.
The enhancement factor $\eta$ is defined as the ratio between the
intensity $I(0)$ and
the ensemble averaged background intensity $\avg{I_\text{bg}}$. This enhancement depends linearly
on the number of degrees of freedom in the wave front.\cite{Vellekoop2007aa}
The value of $\avg{\abs{\gamma(\Delta\mathbf{r})}^2}$ accounts for the loss in overlap between
the sample and the optical fingerprint under sample displacement.

When the range of complexity in the sample is on a subwavelength scale, the
overlap of the optical fingerprint with the sample depends solely on the illumination optics.
In our experiment the pixels of the wave front synthesizer
are projected onto the back aperture
of an infinity corrected microscope objective. In this geometry, we calculated the overlap
for an in-plane displacement to yield
\begin{align}
    \avg{\abs{\gamma(\Delta\mathbf{r})}^2} = \left[
    \frac{2J_1(k_\text{max} \abs{\Delta\mathbf{r}})}
    {k_\text{max} \abs{\Delta\mathbf{r}}}\right]^2,
    \label{eq:gamma2displacement}
\end{align}
where the maximum contributing transversal wave vector $k_\text{max}$ is
determined by the numerical aperture $\text{NA}$ of the microscope objective,
$k_\text{max} = 2\pi \text{NA} / \lambda$. This overlap only
equals unity for $\Delta\mathbf{r} = 0$ and becomes smaller for any
nonzero displacement.

The highest sensitivity of the systems is found by maximizing
the gradient $\nabla$ of the difference intensity
$\Delta I \equiv \Delta I_B - \Delta I_A$.
Using Eq.~\ref{eq:IntensityVsDisplacement} and Eq.~\ref{eq:gamma2displacement}
we find that maximum value of this gradient lies exactly in between $\mathbf{r}_A$
and $\mathbf{r}_B$ when their distance is set to
$\abs{\mathbf{r}_A - \mathbf{r}_B}_\text{opt} = 2.976 / k_\text{max}$.
For these conditions the resulting optimal sensitivity $S_\text{opt}$ is
\begin{align}
\label{eq:sensitivityPositionDetection}
    S_\text{opt} \equiv
    \text{max} \left[
    \nabla (\Delta I)
    \right]
    = \frac{ 5.8 \text{NA} \eta }{\lambda} \avg{I_\text{bg}},
\end{align}
By changing the enhancement $\eta$, the wavelength $\lambda$, and the NA of the optics it is possible to tune the sensitivity over a wide range.

To test our position detection method we employ a spatial light modulator (SLM) from Holoeye (LC-R $2500$) that allows us to spatially modulate the phase of a beam from a continuous wave laser
(Coherent Compass M$315$-$100$, $\lambda = 532$~nm). The spatially modulated beam is then imaged
onto the back aperture of a microscope objective (NA $= 0.95$). In the focal plane of the
microscope objective we have placed a strongly scattering sample on top of a
high precision positioning xyz-stage (Physik Instrumente P-611.3S NanoCube).
The sample is composed of zinc oxide powder on top of a glass cover slide.
At the back of the sample we collect
the transmitted light and image the far field onto a CCD camera (Dolphin F-145B). Because our
method does not need a spatially imaging detector, the camera can be replaced with two photo diodes to maximize bandwidth.

The optimal distance between optimization positions
$\mathbf{r}_A$ and $\mathbf{r}_B$
is calculated to be $252$~nm for this system. Without loss of generality, we consider
only translations in the x-direction. We define the original sample position
as $x_0 = 0$~nm. The sample is translated towards $x_A = -126$~nm.
A feedback based algorithm finds the optical fingerprint for which the scattered
light is focussed on the left side of the camera.
Then we position the sample at $x_B = +126$~nm and repeat the procedure
to create a focus on the right side of the camera. The two
fingerprints are superimposed on the SLM. When we move the sample
back to the original position $x_0$, the two spots become visible on the camera.

\begin{figure}[htb]
\centerline{\includegraphics[width=7.5cm]{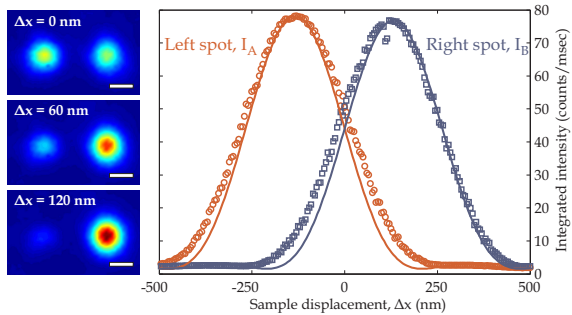}}
  \caption{
  Spot intensities as function of the sample displacement.
  On the left there are three camera images for different
  values of the sample displacement $\Delta x$. As $\Delta x$ increases
  the intensity $I_A$ in the left spot decreases while the intensity $I_B$
  in the right spot increases. The scale bars denote $1$~mm.
  The graph on the right shows the measured spot intensities (circles $I_A$,
  squares $I_B$) as function of the sample displacement. The solid lines denote
  the theoretical expected behavior.
  }
  \label{fig:vanPutten2011_Fig3}
\end{figure}
In Fig.~\ref{fig:vanPutten2011_Fig3} the intensity in the two spots
as a function of the sample displacement $\Delta x \equiv x - x_0$ is plotted together
with camera images for three different values of $\Delta x$. The two lines
denote the intensity behavior predicted by Eq.~\ref{eq:IntensityVsDisplacement}
and Eq.~\ref{eq:gamma2displacement} without free parameters.
Starting from $ \Delta x = 0$ where the intensity in
both spots is equal, $I_A$ and $I_B$ change differently under sample displacement.
Moving the sample in the positive x-direction results in a decreasing $I_A$ while
$I_B$ increases for $\Delta x < x_B$. The experimental data is in good agreement
with theory although the measured intensity dependence is slightly wider. This
small deviation is likely to be caused by a non-ideal transfer function of the optics, which cannot be compensated for by wave front corrections.

\begin{figure}[htb]
\centerline{\includegraphics[width=7.5cm]{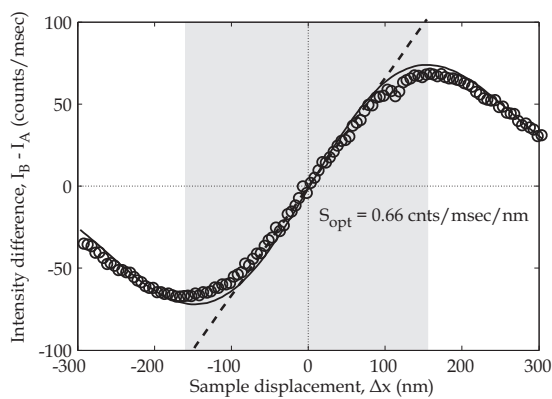}}
  \caption{
  Intensity difference $I_B - I_A$ as function of the sample displacement $\Delta x$.
  The circles denote the measured intensity difference, while the solid line
  represents the theoretical intensity difference. Within the gray area the
  function is bijective. The dotted line is a linear fit to the data points close
  to $\Delta x = 0$. From the fit we find that the slope equals
  $S_\text{opt} = 0.66$~counts/msec/nm.
  }
  \label{fig:vanPutten2011_Fig4}
\end{figure}
To find the position of the sample from the measured data we look at the
difference intensity between the two spots, which is plotted in
Fig.~\ref{fig:vanPutten2011_Fig4}. For $x_A > \Delta x > x_B$ (gray area)
the function is bijective resulting in a unique mapping between the difference signal
and the sample position. Over a large distance, the function is linear and
only close to the displacements $\Delta x = x_A$ and $\Delta x = x_B$
the function starts to curve. The highest sensitivity is found at
$\Delta x = 0$ where the slope has a maximum value of $S_\text{opt} = 0.66$~counts/msec/nm,
close to the theoretical limit for this system of $0.80$~counts/msec/nm that we calculated using
Eq.~\ref{eq:sensitivityPositionDetection}.
The noise level in our setup is found to be $1.42$~counts/msec, so that the achievable
displacement resolution is $2.1$~nm. We know that part of the noise may infact be signal, i.e., fluctuations of the actual sample position.
The achieved resolution compares favorably with state of the art techniques\cite{Pedrini2011_OptEng}. A higher resolution is possible by
increasing the signal-to-noise ratio in the system by, e.g., increasing the
intensity enhancement $\eta$ in the spots.

Instead of measuring the scattered light in transmission, one could also choose to work in reflection. Furthermore, by employing more than two detectors and generating multiple optical fingerprints,
the method can be expanded in a straightforward way to simultaneously detect displacements
in multiple directions. Similarly, the optical fingerprints can also be configured to
detect other sample movements, such as rotations. This flexibility makes our technique
very suitable for high speed and high precision position monitoring of complex scattering structures.

The authors thank I.M. Vellekoop and W.L. Vos for their valuable support.
I.D. Setija from ASML Netherlands B.V. is acknowledged for stimulating discussions.
A.P. Mosk is supported by a Vidi grant from NWO.

\end{document}